\documentclass[12pt]{article}
\usepackage{graphicx} % Required for inserting images
\usepackage{amssymb,amsmath}
\usepackage{todonotes}
\usepackage{fullpage}
\usepackage{url}
\usepackage{color}
\definecolor{colorS}{RGB}{0,154,128}
\definecolor{colorI}{RGB}{255,32,0}
\definecolor{colorR}{RGB}{205,10,179}
\definecolor{colorE}{RGB}{240,228,66} 

\usepackage{tikz}
\usetikzlibrary{arrows.meta,fit,backgrounds}
\tikzstyle{transition}=[rectangle]

\tikzset{
    state/.style={
        draw, rounded corners=2pt,
        minimum width=18mm, minimum height=10mm,
        align=center
    },
    phi/.style={
        state, fill=colorS!35
    },
    1minustheta/.style={
        state, fill=colorI!35
    },
    thetafit/.style={
        draw, dashed, rounded corners=4pt,
        fill=colorS!15
    },
    Sstatus/.style={
        state, fill=colorS!25
    },
    Estatus/.style={
        state, fill=colorE!25
    },
    Istatus/.style={
        state, fill=colorI!25
    },
    Rstatus/.style={
        state, fill=colorR!25
    },
    flow/.style={
        -{Latex[length=2mm]}, thick
    },
    rate/.style={
        font=\small, inner sep=2pt
    }
}

\title{Low Dimensional Exact Edge-Based Compartmental Modeling for Simple Contagions}
\author{Joel C. Miller}

\newcommand{\ave}[1]{\left \langle #1 \right \rangle}
\newcommand{\status}[1]{\ensuremath{\mathrm{#1}}}
\newcommand{\Q}{\status{Q}}
\newcommand{\I}{\status{I}}
\newcommand{\E}{\status{E}}
\newcommand{\R}{\status{R}}
\newcommand{\Sus}{\status{S}}

\begin{document}
\maketitle
\begin{abstract}
    The Edge-Based Compartmental Model (EBCM) first introduced for the spread of Susceptible-Infected-Recovered (SIR) disease in static networks provides a low-dimensional system of equations which exactly predict the stochastic dynamics in the large population limit.  We show that the approach can be adapted for a class of simple contagions (that is, a process in which nodes may change status due to independent interactions with neighbors).  If nodes can begin susceptible but never transition into a susceptible state and the initial node statuses are independently assigned, then it is possible to derive EBCM equations for the simple contagion spreading in a Configuration Model network.  We derive the relevant equations and numerically demonstrate their accuracy for several simple contagions using automated algorithms in the Epidemics on Networks Python package.
\end{abstract}
\section{Introduction}

The study of contagious processes (or \emph{contagions}) in networks has stimulated a significant amount of research.  Although much of the original research has focused on infectious diseases spreading through contact networks~\cite{kiss2017mathematics,porter2016dynamical,pastor2015epidemic}, more recently these concepts have been applied more broadly to the spread of other contagions through networks.  Examples include rumors, beliefs, or cultural practices that can also be thought of as spreading contagiously through social networks~\cite{castellano2009statistical}.

In the context of social media or politics, contagions play an important role in polarization~\cite{bail2018exposure}.  They drive many political movements as well as other social processes.  Their effects can be positive or negative.  They can be used to spread valuable public health messages, or to promote misinformation~\cite{vosoughi2018spread}. They can be exploited by hostile nations~\cite{bradshaw2019global} or used by social media companies to attract attention and increase engagement~\cite{milli2025engagement}.

Broadly, we will focus on contagions spreading on networks as processes with the following assumed properties:
\begin{itemize}
    \item Each node has a discrete \emph{status}.\footnote{We recommend that researchers avoid using the word ``state'' for the status of a given node so that we can reserve ``state'' to refer to the state of the entire network without confusion.}
    \item A node may \emph{transition} from one status to another as an instantaneous event in response to the statuses of its neighbors or as a spontaneous event that is not affected by its neighbors.  
\end{itemize}
In this study, we will focus on Poisson processes such that a node $u$ with status $\Q_x$ changes to status $\Q_z$ with a rate $f_{x,z}(n_1, n_2, \ldots, n_m)$ where $n_y$ is the number of neighbors of $u$ with status $\Q_y$.  

One of the best-studied contagions is the \emph{Susceptible-Infected-Recovered} (SIR)  model.  In this model, most nodes begin susceptible (\Sus).  If a neighbor is infected (\I), a susceptible node may transition from \Sus{} to \I{} as a Poisson process with rate $\tau$.  If multiple neighbors are infected, it transitions with rate $\tau n_\I$ where $n_\I$ is the number of infectious neighbors.  This is equivalent to each neighbor independently causing a transition with rate $\tau$.  Infected nodes spontaneously recover with rate $\gamma$.  Whether a node recovers in a short time interval is independent of how many infected or susceptible neighbors it might have.  

It is common to think of two classes of contagion, \emph{simple contagions} and \emph{complex contagions}~\cite{centola2007complex,dodds2004universal,guilbeault2018complex}. 

For simple contagions (such as the SIR model), an individual's status changes in response to individual neighbors or spontaneously due to intrinsic properties of itself.  The neighbors can be treated as acting independently.  Assuming that transitions occur as a Poisson process, the rate that a node with status $\Q_x$ transitions to status $\Q_z$ is given by
    \begin{align}
    \lambda_{x,z} &= f_{x,z}(n_1, n_2, \ldots, n_m) \nonumber\\
    &= a_{x,z} + \sum_{y =1}^m   b_{x,z}^{(y)} n_{y}
    \label{eqn:simple_contagion_def}
    \end{align}
where $a_{x,z}\geq 0$ is the rate that a status $\Q_x$ individual would spontaneously transition to type $\Q_z$ in isolation, while $b_{x,z}^{(y)} \geq 0$ is the rate at which a status $\Q_x$ individual would transition to $\Q_z$ due to interactions with a single neighbor of status  $\Q_y$.  The independence of neighbors manifests itself in the fact that the transition rate is the spontaneous rate $a_{x,z}$ plus the linear combination $\sum_{y=1}^m b_{x,z}^{(y)} n_y$~\cite{kiss2017mathematics}.

For complex contagions, an individual's status changes due to its neighbors in such a way that the neighbors do not act independently.  That is, $f_{x,z}$ does not have the simple form of Equation~\eqref{eqn:simple_contagion_def}.  In practice, this lack of independence could represent individuals becoming much more likely (or less likely) to believe a claim the second time that they have heard it.

In this paper, we consider the spread of simple contagions.  We will assume that the underlying population structure corresponds to a Configuration Model network~\cite{molloy1995critical,newman2001random}.  In a Configuration Model network, each node $u$ is assigned a degree $k_u$ following some probability distribution.  We assign $k_u$ ``stubs'' or half-edges to the node $u$.  Once all nodes have stubs, we select pairs of stubs randomly and join them into edges.  

In principle, if we ever create two edges between the same pair of nodes or a self-edge, we should start over.  In practice under weak assumptions on the degree distribution~\cite{janson2009probability}, the number of such edges remains small as the number of nodes $N \to \infty$: Asymptotically, the number of self-edges has a Poisson distribution with mean $\frac{\langle K^2\rangle - \langle K \rangle }{2\langle K \rangle}$ where $\langle K^2 \rangle$ and $\langle K \rangle$ are the average squared degree and the average degree respectively.  Similarly, the number of pairs of repeated edges is asymptotically Poisson, but with mean equal to the square: $\left(\frac{\langle K^2\rangle - \langle K \rangle}{2\langle K \rangle}\right)^2$~\cite{janson2009probability}.  So in the $N\to\infty$ limit the probability that any node is in one of these edges, or is even within a short distance of one of these edges is negligible.

\subsection{Deterministic modeling techniques}

We briefly review some analytic methods that have been widely used to model infectious disease spread in Configuration Model networks in the large $N$ limit.  We focus particularly on how they are used to model the spread of SIR disease.  More complete details about these models are given in~\cite{kiss2017mathematics,miller2014epidemic}.

\subsubsection{Pairwise Models}
In Pairwise Models~\cite{keeling1997correlation,keeling1999effects,eames2002modeling,house2011insights,miller2014epidemic,kiss2017mathematics}, we track variables such as $[S_k]$, $[I_k]$, and $[R_k]$ which count the number of nodes of each degree and of each status.  

This already requires $\mathcal{O}(K)$ variables, where $K$ is the number of distinct degrees.  The rate of change of quantities such as $[S_k]$ depends on knowing how many pairs there are which include a susceptible degree-$k$ node joined to an infected node.  Often researchers build a \emph{heterogeneous mean field model}~\cite{pastor2001epidemic,moreno2002epidemic,pastor2015epidemic,kiss2017mathematics} which implicitly assumes that the statuses of nodes and their neighbors are independent (or equivalently that nodes reselect their neighbors at each moment in time).  However, because the disease spreads along edges, this independence assumption violates the infectious process if a population has long-lasting partnerships.  Under our assumption of a static network, an infected node's neighbors are more likely to be infected than the neighbors of a susceptible node.

To avoid this inaccurate assumption, we introduce new variables for pairs, such as $[S_{k_1} I_{k_2}]$ which counts the number of edges that connect a susceptible degree $k_1$ node with an infected degree $k_2$ node.  This increases the number of variables to $\mathcal{O}(K^2)$.  

In principle, we need variables involving triples to write down equations for the pairs, which would lead then to $\mathcal{O}(K^3)$.  However, we can close the equations by writing the triples in terms of the pairs, keeping the complexity at $\mathcal{O}(K^2)$ equations.   The resulting model is known as a \emph{pair approximation} or a \emph{pairwise model}.  In the case of the SIR model these equations are exact (in the limit of large Configuration Model networks, subject to mild conditions~\cite{decreusefond2012large, janson2014law,miller2014epidemic,kiss2017mathematics}).   For SIS disease, in which recovered nodes are susceptible again, the pairwise model results in approximate equations~\cite{miller2014epidemic,kiss2017mathematics}.

\subsubsection{Effective Degree Models}
The Effective Degree Models~\cite{lindquist2011effective,gleeson2011high,miller2014epidemic,kiss2017mathematics} focus on variables such as $S_{s,i,r}$, $I_{s,i,r}$, and $R_{s,i,r}$.  Here $S_{s,i,r}$ counts the number of susceptible nodes with $s$, $i$, or $r$ neighbors who are susceptible, infected, or recovered respectively.  We similarly define $I_{s,i,r}$ and $R_{s,i,r}$.  Through various simplifications, it is possible to reduce this to $\mathcal{O}(M^2)$ equations where $M$ is the maximum degree.  

The Effective Degree model is precise in the large Configuration Model network limit for SIR disease, but not for SIS disease~\cite{decreusefond2012large,janson2014law,miller2014epidemic,kiss2017mathematics}.  For SIS dynamics the Effective Degree Model usually outperforms the Pairwise Model~\cite{kiss2026mathematics}.

\subsubsection{Edge-based Compartmental Models}
The Edge-Based compartmental modeling (EBCM) approach focuses on whether any edges of a node have transmitted to it or not~\cite{volz2008sir,miller2011note,miller2011edge}.  These equations are used for SIR-like diseases for which an individual can only be infected once.  Under the assumption that the initial node statuses are all assigned independently, the pairwise and effective degree models can be systematically reduced to the EBCM equations~\cite{miller2014epidemic,kiss2017mathematics,ibrahim2022generating}.  Under relatively broad conditions on the degree distribution, the EBCM model is precise for SIR dynamics on a static Configuration Model network~\cite{decreusefond2012large,janson2014law}.

The EBCM approach uses a probability generating function (PGF)~\cite{wilf2014generatingfunctionology,newman2001random,miller2018primer} to encode the degree distribution of the underlying network.  It introduces a variable $\theta(t)$ which measures the probability that an edge has not yet transmitted infection to the node.  If $\psi(x)$ is the PGF of the degree distribution, then $\psi(\theta(t))$ can be used to concisely express the probability that a random node is susceptible~\cite{volz2008sir,miller2011note,miller2011edge}.  This allows us to reduce the number of equations to $\mathcal{O}(1)$ (in contrast, the pairwise and effective degree models have infinitely many equations if the maximum degree is unbounded).  

For the standard SIR case, there is a single differential equation for $\theta(t)$.  After solving that equation, $S(t)$ can be found through an algebraic expression.  Using an additional differential equation and conservation rules we can find $I(t)$ and $R(t)$.

Two key facts underlie the EBCM approach for SIR disease in a Configuration Model network.  We illustrate these by considering a node $u$ and two of its (possibly many) neighbors $v$ and $w$ in the limit of a large Configuration Model network.  
\begin{itemize}
    \item If we know that $u$ is susceptible at time $t$, then knowing the status of $v$ does not give any information about the status of $w$.  That is, conditional on $u$ being susceptible, the statuses of $v$ and $w$ are independent (In contrast, if $u$ is infected, knowing the status of $v$ gives information about how long $u$ has been infected which in turn gives information about the possible status of $w$).
    \item Once $u$ becomes infected, it will eventually recover spontaneously, after which it can never become infected again.  The duration of its infection is not impacted by $v$ or $w$ (or any other neighbor).  
\end{itemize}
Thus, to calculate the probability that $u$ is still susceptible, we can treat $v$ and $w$ as independent.  Once $u$ becomes infected, the statuses of $v$ and $w$ have no further effect on $u$'s status.  Thus we can treat neighbors of $u$ as independent in all of our calculations of $u$'s probability of having any given status.

The two key facts listed above also underlie why the pairwise and effective degree models are exact for SIR dynamics in large Configuration Model networks.  The EBCM approach has been extended to a wide range of models, including live oral polio vaccine~\cite{balde2019contact}, clustered networks~\cite{barnard2018edge,volz2011effects}, complex contagions~\cite{wang2015dynamics,miller2016complex}, and misinformation spread~\cite{xian2019misinformation}.

\subsection{Overview}
In this paper we will derive EBCM equations for a class of simple contagions spreading in Configuration Model networks.  To help define the specific class of simple contagions, we define 
\begin{itemize}
    \item \emph{susceptible statuses} to be those statuses which can be changed due to interactions with neighbors. 
    \item \emph{infectious statuses} to be those statuses which may cause neighbor statuses to change.
\end{itemize}
Note that these definitions are not exclusive: a node may be simultaneously susceptible and infectious.  It was conjectured in~\cite{kiss2026mathematics} that given a simple contagion for which
\begin{itemize}
         \item it is not possible to transition into any susceptible status
         \item and no susceptible status is also infectious
\end{itemize}
With a constraint on the initial condition such that
\begin{itemize}
         \item each node's initial status is assigned independently of the others
\end{itemize}
then it is possible to write down an EBCM model which accurately captures the dynamics in the large population limit (subject to regularity conditions on the network degree distribution).

We will develop a stronger result, dropping the second condition so that susceptible nodes may also be infectious.  We derive the EBCM equations for arbitrary simple contagions for which it is not possible to transition into a susceptible status and the initial statuses are assigned independently, such that the probability of status $\Q_x$ is $\rho_x$ independently for each node.  Thus given a simple contagion and initial condition for which 
\begin{itemize}
         \item it is not possible to transition into any susceptible status and
         \item each node's initial status is assigned independently of the others
\end{itemize}
then we can use the EBCM approach to derive equations that predict the dynamics of the contagion spreading in a Configuration Model network.  

We do not prove specific conditions under which this model is accurate.  We would expect that if the initial numbers of some classes are small enough stochastic effects may impact the long-term dynamics.  Similarly, we anticipate that if the tail of the degree distribution decays too slowly, the highest degree individual nodes in the population could have a population-scale impact, and thus the deterministic model will not apply.  However, we expect this model to apply across a wide range of initial conditions and degree distributions.

\section{Derivation}
We consider a simple contagion with $m$ statuses $\Q_1$, $\Q_2$, $\ldots$, $\Q_m$.  A node of status $\Q_x$ with $n_y$ neighbors of status $\Q_y$ transitions from status $\Q_x$ to $\Q_z$ with rate 
\[
\lambda_{x,z} = a_{x,z} + \sum_{y=1}^m b_{x,z}^{(y)}n_y
\]
The $a_{x,z}$ terms correspond to spontaneous transitions due to intrinsic properties of the individual inducing a transition from status $\Q_x$ to $\Q_z$.  An example of this occurs in the SIR or SIS models when an individual recovers from infection.

The $b_{x,z}^{(y)}$ terms correspond to transitions due to interactions with individual neighbors of status $\Q_y$ inducing a change from $\Q_x$ to $\Q_z$.  An example of this occurs in the SEIR model when an infected node can cause a susceptible node to transition to the exposed state.  We borrow from the language of infectious disease saying that a status $\Q_y$ neighbor \emph{transmits} to a status $\Q_x$ node when it induces the change in status from $\Q_x$ to $\Q_z$.  The transmission occurs with rate $b_{x,z}^{(y)}$ along each $\Q_x$ $\Q_y$ edge.  

\subsection{Graphical Representation of Simple Contagions}
We create two new directed \emph{transition graphs} $\mathcal{H}$ and $\mathcal{J}$.  They define the potential transitions nodes can undergo and can be used to describe the contagious process.  They have nothing to do with the underlying structure of the contact network.

The directed graph $\mathcal{H}$ represents the \emph{spontaneous transitions} that a node can undergo without interaction with its neighbors.  The nodes of $\mathcal{H}$ are the possible statuses $\Q_1, \ldots, \Q_m$.  If $a_{x,z}>0$, then a weighted edge $\Q_x \mapsto \Q_z$ exists with weight $a_{x,z}$.  Thus an edge in $\mathcal{H}$ represents a spontaneous transition that is possible in the simple contagion and its weight represents the rate of that transition. We typically only include nodes in $\mathcal{H}$ if their in- or out-degree is positive.

The directed graph $\mathcal{J}$ represents the neighbor-induced transitions.  These correspond to transmissions from neighbors that cause a node to change status.  The nodes of graph $\mathcal{J}$ are pairs of statuses of the form $(\Q_y, \Q_x)$ and an edge in $\mathcal{J}$ takes the form $(\Q_y, \Q_x) \mapsto (\Q_y, \Q_z)$ to represent that a node of status $\Q_y$ can induce a transition from status $\Q_x$ to $\Q_z$.  The corresponding weight of the edge is $b_{x,z}^{(y)}>0$ and gives the rate of the transition.  As with $\mathcal{H}$ we typically only include nodes in $\mathcal{J}$ if their in- or out-degree is positive.

Figure~\ref{fig:SIRS} illustrates the graphs $\mathcal{H}$ and $\mathcal{J}$ for an SIRS disease: susceptible nodes that interact with infected nodes may become infected (given by the $(\I,\Sus) \mapsto (\I,\I)$ edge in $\mathcal{J}$), once infected, nodes spontaneously recover (given by the $\I \mapsto \R$ edge in $\mathcal{H}$) and later spontaneously return to a susceptible state (given by the $\R \mapsto \Sus$ edge in $\mathcal{H}$).

\begin{figure}
\centering
    \framebox{
     \begin{tikzpicture}
                \node[transition] (I) at ( 1,0) {\I};
        \node[transition] (R) at ( 3,0) {\R};
        \node[transition] (S) at ( 5,0) {\Sus};

        \draw [->,sloped, above] (I)--  node {$\gamma$} (R);
        \draw [->,sloped, above] (R)--  node {$\omega$} (S);
        \node [] (H) at (0, 0) { $\mathcal{H}$:};
        \node[transition] (IS) at ( 9,0) {$(\I,\Sus)$};
        \node[transition] (II) at ( 12,0) {$(\I, \I)$};
        \draw [->,sloped, above] (IS)--  node {$\tau$} (II);
        \node [] (J) at (7.5, 0) {$\mathcal{J}$:};
    \end{tikzpicture}}
\caption{The transition graphs $\mathcal{H}$ and $\mathcal{J}$ for the SIRS model.  Susceptible nodes become infected due to interactions with infected neighbors.  Infected nodes spontaneously recover and later spontaneously regain susceptibility.  This contagion does not satisfy the assumptions we require for an EBCM model to exist as nodes eventually return to a susceptible status after recovering.}
\label{fig:SIRS}
\end{figure}

We will say that a node with status $\Q_x$ is \emph{susceptible} if there is some status $\Q_y$ such that a $\Q_y$ node can transmit to a $\Q_x$ node.  That is, a node is susceptible if there is at least one edge in $\mathcal{J}$ of the form $(\Q_y,\Q_x) \mapsto (\Q_y, \Q_z)$ for some $y$ and $z$.  Otherwise we will say that a node with status $\Q_x$ is not susceptible.  If it is not susceptible, then a node with that status can only undergo spontaneous transitions.  For example, in the standard SIR model, the status $S$ is susceptible while both $I$ and $R$ are not.

Note that under our definition, it is possible for a node with a susceptible status to transmit to one of its neighbors.  Thus unlike traditional infectious disease modeling, ``susceptible'' and ``infectious'' are not necessarily mutually exclusive.

\begin{figure}
    \centering
    \framebox{
    \begin{tikzpicture}
        \node[transition] (E) at ( 1,0) {$\status{E}$};
        \node[transition] (I) at ( 3,0) {$\I$};
        \node[transition] (R) at ( 5,0) {$\R$};

        \draw [->,sloped, above] (E)--  node {$\sigma$} (I);
        \draw [->,sloped, above] (I)--  node {$\gamma$} (R);
        \node [] (H) at (0, 0) { $\mathcal{H}$:};
            \node[transition] (IS) at ( 9,0) {$(\I,\Sus)$};
        \node[transition] (IE) at ( 12,0) {$(\I, \status{E})$};
        \draw [->,sloped, above] (IS)--  node {$\tau$} (IE);
        \node [] (J) at (7.5, 0) {$\mathcal{J}$:};

        \node [] (a) at (0,-1) {\textbf{(a)}};
    \end{tikzpicture}}

    \vspace{24pt}

\framebox{
    \begin{tikzpicture}
        \node[transition] (S) at (1,2) {$\Sus$};
        \node[transition] (I1) at ( 3,2) {$\I_1$};
        \node[transition] (I2) at ( 3,1) {$\I_2$};
        \node[transition] (R1) at ( 6,2) {$\R_1$};
        \node[transition] (R2) at ( 6,1) {$\R_2$};

        \draw [->, sloped, above] (S) -- node {$\omega$} (I1);
        \draw [->,sloped, above] (I1)--  node {$\gamma_1$} (R1);
        \draw [->,sloped, below] (I2)--  node {$\gamma_2$} (R2);
        \node [] (H) at (0, 2) { $\mathcal{H}$:};
        \node[transition] (I1S) at (9,1.5) {$(\I_1,\Sus)$};
        \node[transition] (I1I1) at ( 12,2) {$(\I_1, \I_1)$};
        \node[transition] (I1I2) at ( 12,1) {$(\I_1, \I_2)$};
        \node[transition] (I2S) at (9,0) {$(\I_2,\Sus)$};
        \node[transition] (I2I2) at ( 12,0) {$(\I_2, \I_2)$};
        \draw [->,sloped, above] (I1S)--  node {$p\tau_1$} (I1I1);
        \draw [->,sloped, below] (I1S)--  node {$(1-p)\tau_1$} (I1I2);
        \draw [->,sloped, below] (I2S)--  node {$\tau_2$} (I2I2);
        
        \node [] (J) at (7.5, 2) {$\mathcal{J}$:};

        \node [] (b) at (0,0) {\textbf{(b)}};
    \end{tikzpicture}}

    \vspace{24pt}

% \framebox{
%     \begin{tikzpicture}
%         \node[transition] (I1) at ( 1,2) {$\I_1$};
%         \node[transition] (I2) at ( 1,0) {$\I_2$};
%         \node[transition] (R) at ( 4,1) {$\R$};
%         %

%         \draw [->,sloped, above] (I1)--  node {$\gamma_1$} (R);
%         \draw [->,sloped, below] (I2)--  node {$\gamma_2$} (R);
%         \node [] (H) at (0, 2) { $\mathcal{H}$:};
% %
% %
%         \node[transition] (I1S) at (9,1.5) {$(\I_1,\Sus)$};
%         \node[transition] (I1I1) at ( 12,2) {$(\I_1, \I_1)$};
%         \node[transition] (I1I2) at ( 12,1) {$(\I_1, \I_2)$};
% %
%         \node[transition] (I2S) at (9,0) {$(\I_2,\Sus)$};
%         \node[transition] (I2I2) at ( 12,0) {$(\I_2, \I_2)$};
% %
%         \node[transition] (SS) at (9, -1) {$(\Sus,\Sus)$};
%         \node[transition] (SI1) at (12,-1) {$(\Sus,I_1)$};   
% %
%         \draw [->,sloped, above] (I1S)--  node {$p\tau_1$} (I1I1);
%         \draw [->,sloped, below] (I1S)--  node {$(1-p)\tau_1$} (I1I2);
%         \draw [->,sloped, below] (I2S)--  node {$\tau_2$} (I2I2);
%         \draw [->, sloped, below] (SS)-- node {$\tau_3$} (SI1);
% %
%         \node [] (J) at (7.5, 2) {$\mathcal{J}$:};
%     \end{tikzpicture}}

\framebox{
\begin{tikzpicture}
    % Graph H
    \node[transition] (I1) at (1,0) {$\I_1$};
    \node[transition] (I2) at (1,-2) {$\I_2$};
    \node[transition] (R) at (4,-1) {$\R$};

    \draw [->,sloped,above] (I1)-- node {$\gamma_1$} (R);
    \draw [->,sloped,below] (I2)-- node {$\gamma_2$} (R);
    \node (H) at (0,0) {$\mathcal{H}$:};

    % Graph J
    \node[transition] (I1S1) at (9,0) {$(\I_1,\Sus_1)$};
    \node[transition] (I1S2) at (9,-1) {$(\I_1,\Sus_2)$};
    \node[transition] (I1I1) at (12,-0.5) {$(\I_1,\I_1)$};

    \node[transition] (I2S1) at (9,-2) {$(\I_2,\Sus_1)$};
    \node[transition] (I2S2) at (9,-3) {$(\I_2,\Sus_2)$};
    \node[transition] (I2I2) at (12,-2.5) {$(\I_2,\I_2)$};

    \node[transition] (S1S2) at (9,-4) {$(\Sus_1,\Sus_2)$};
    \node[transition] (S1I2) at (12,-4) {$(\Sus_1,\I_2)$};

    \node[transition] (S2S1) at (9,-5) {$(\Sus_2,\Sus_1)$};
    \node[transition] (S2I1) at (12,-5) {$(\Sus_2,\I_1)$};

    \draw [->,sloped,above] (I1S1)-- node {$\tau_{11}$} (I1I1);
    \draw [->,sloped,below] (I1S2)-- node {$\tau_{12}$} (I1I1);

    \draw [->,sloped,above] (I2S1)-- node {$\tau_{21}$} (I2I2);
    \draw [->,sloped,below] (I2S2)-- node {$\tau_{22}$} (I2I2);

    \draw [->,sloped,above] (S1S2)-- node {$\alpha$} (S1I2);
    \draw [->,sloped,above] (S2S1)-- node {$\beta$} (S2I1);

    \node (J) at (7.5,0) {$\mathcal{J}$:};
    \node [] (c) at (0,-5) {\textbf{(c)}};
\end{tikzpicture}}
    \vspace{24pt}

    \caption{Transition graphs for different simple contagions.  \textbf{(a)}: An SEIR model.  $\mathcal{J}$ shows that susceptible individuals can transition to exposed by interactions with an infectious individual.  Once exposed, $\mathcal{H}$ shows the spontaneous transitions to infected and then recovered occur independently of interactions with neighbors.  \textbf{(b)}:
    an SI$_1$I$_2$R$_1$R$_2$ model.  In this model, there are two types of possible infections.  $\mathcal{H}$ shows that a susceptible individual may spontaneously transition into status $\I_1$ and  $\I_1$ and $\I_2$  recover into $\R_1$ and $\R_2$.  $\mathcal{J}$ shows that those of status $\I_1$ can transmit to susceptible nodes, causing them to transition into either type $\I_1$ (with probability $p$) or $\I_2$ (with probability $1-p$).  Type $\I_2$ nodes can cause susceptible nodes to transition to type $I_2$. \textbf{(c)}: An S$_1$S$_2$I$_1$I$_2$R model with two susceptible and two infectious states in which the susceptible states can transmit to one another.}
    \label{fig:simple_contagion_HJ}
    \end{figure}
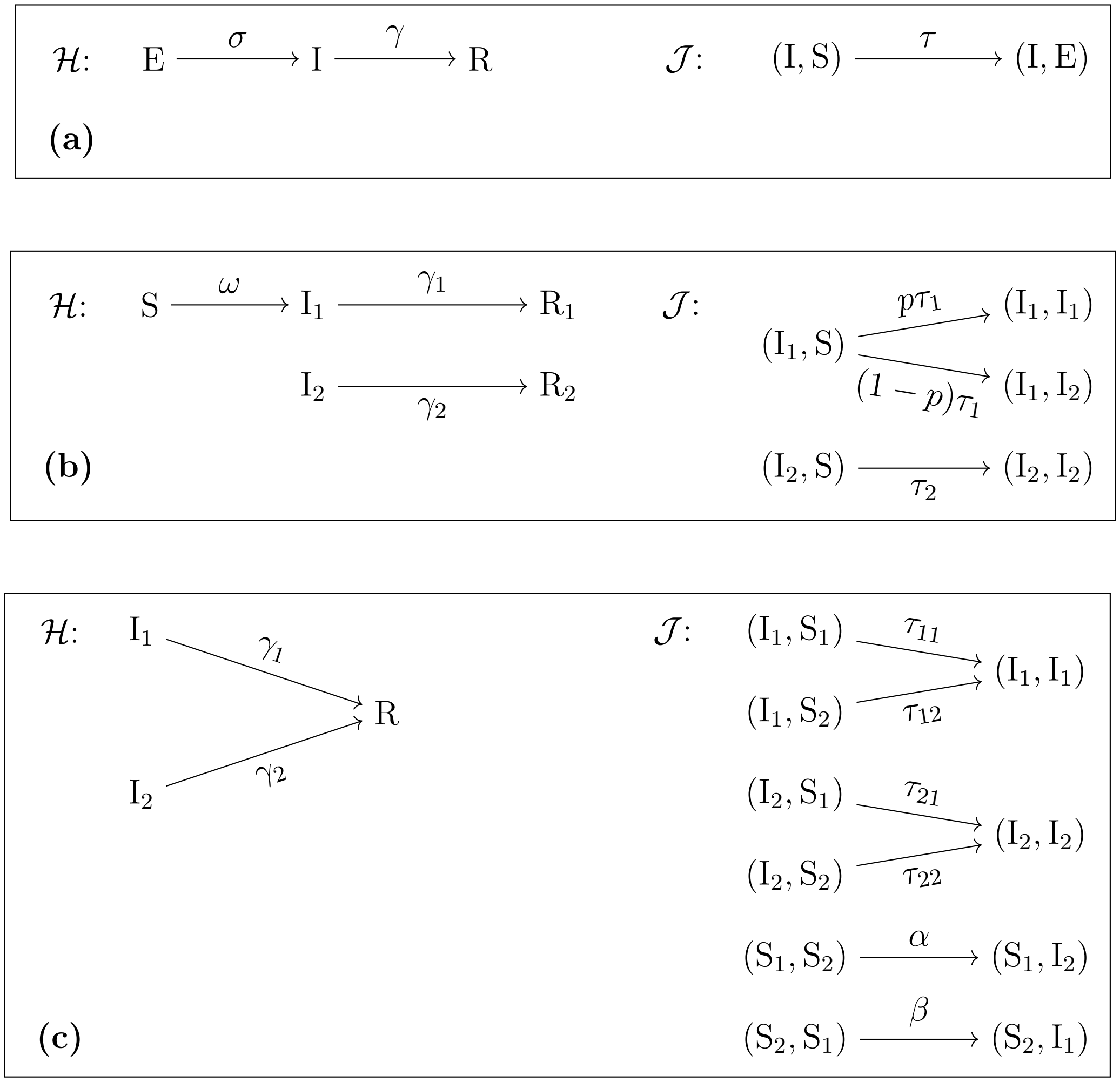

We can assume that the susceptible statuses are $\Q_1, \Q_2, \ldots, \Q_\ell$ and that the remaining statuses $\Q_{\ell+1}, \ldots, \Q_m$ are not susceptible.  To derive an EBCM model, we will focus on the subset of simple contagions for which nodes can never transition into a susceptible status.  They may begin susceptible, but once they undergo any transition, they will never be susceptible again.  The simple contagions shown in Fig~\ref{fig:simple_contagion_HJ} satisfy this property, but the SIRS model in Fig~\ref{fig:SIRS} does not.

\subsection{Equation Derivation}
We assume that at time $0$, statuses are assigned independently to each node, each with probability $\rho_y$.  We define $Q_y(t)$ to be the proportion of the network to have status $\Q_y$ at time $t$.  So $Q_y(0) = \rho_y$.  

We make the assumption that the number of nodes in the network $N$ is large enough, and $N\rho_y$ are all large enough that stochastic effects are negligible (we will see that in some cases we can ignore stochastic effects even if some statuses are initially absent).  Under this assumption, we are able to calculate the proportion of nodes having each status $Q_1(t)$, $Q_2(t)$, $\ldots$, $Q_m(t)$ by instead calculating the probability a random node will have each status at time $t$ given the initial conditions.

\begin{figure}
\centering
    \begin{tikzpicture}
    \node[Sstatus] (S) at (0,0) {$S$};
    \node[Istatus] (I) at (4,0) {$I$};
    \node[Rstatus] (R) at (8,0) {$R$};

    \draw[flow] (S) -- node[rate, above]
        {$\beta I S$} (I);

    \draw[flow] (I) -- node[rate,above]
        {$\gamma I$} (R);
\end{tikzpicture}
    \caption{Flow diagram for a traditional mass action model, which ignores network structure and degree heterogeneity.  When we introduce the network structure, the $I$ to $R$ flow remains the same, but the $S$ to $I$ flow will change.}
    \label{fig:MAflow}
\end{figure}
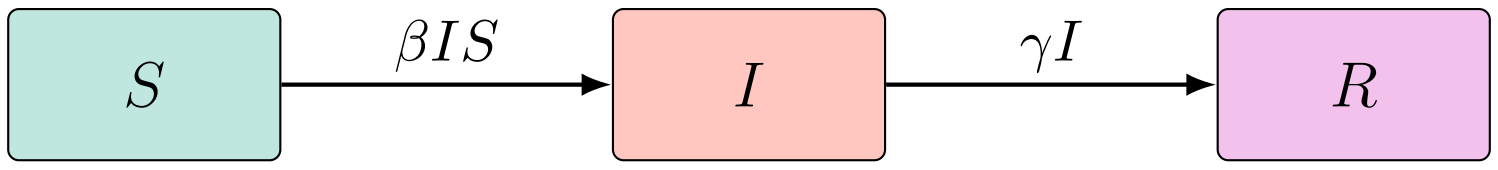

For standard mass action models, we can derive our equations using a flow diagram like in Fig.~\ref{fig:MAflow} which describes the SIR model.  In this model, the flow from $I$ to $R$ is a spontaneous transition.  It does not depend on interactions with neighbors.  However, the flow from $S$ to $I$ involves interactions with neighbors.  

When we try to create a flow diagram that accounts for the network structure, the flow from $I$ to $R$ will remain the same. In general any flows corresponding to spontaneous transitions (those in $\mathcal{H}$) can be captured similarly. However, the flow from $S$ to $I$ will be different.  More generally, any flows that involve interactions with neighbors cannot be captured in the same way as in the mass action models.  Our effort in deriving equations for our models will focus on determining the flows resulting from interactions with neighbors (those in $\mathcal{J}$).

So the flow between non-susceptible statuses  proceeds exactly as in the equivalent mass action models.  However the flow from susceptible statuses changes.  To calculate the flows from susceptible statuses will require a new approach.

\subsubsection{Susceptible statuses}
If $\Q_x$ is a susceptible status, then the probability that a random node $u$ has status $\Q_x$ at time $0$ is $\rho_x$.  We will find that the probability that it still has status $\Q_x$ at time $t$ given that its degree is $k$ is 
\[
\mathbb{P}[\text{status }\Q_x \text{ at time }t | \text{degree }k] = \rho_x e^{-t(\sum_z a_{x,z})}\theta^{(x)}(t)^k
\]
Here $e^{-t(\sum_z a_{x,z})}$ gives the probability that a node $u$ would not have undergone a spontaneous transition (conditional on $u$ not receiving any transmissions from its neighbors).   The variable $\theta^{(x)}(t)$ measures the probability that a random neighbor $v$ of $u$ has not transmitted to $u$ by time $t$ (conditional on $u$ not undergoing a spontaneous transition and no other neighbor transmitting to $u$).  A more precise definition of $\theta^{(x)}$ will be given later.

Taking a weighted sum over all degrees and defining the probability generating function $\psi(s) = \sum_k P(k) s^k$, the probability a randomly chosen node $u$ has a susceptible status $\Q_x$ is 
\[
Q_x(t) = \rho_x e^{-t (\sum_z a_{x,z})} \psi(\theta^{(x)}(t))
\]
If we can determine $\theta^{(x)}(t)$ for each susceptible status and the fluxes from each susceptible status to each non-susceptible status, we will have a complete system.

\subsubsection{Calculation of $\theta^{(x)}(t)$}

In our brief description of $\theta^{(x)}$ above, we assumed that $u$ has not undergone any other transitions by time $t$. If $u$ changes status before $t$ and then causes $v$ to change status, this would itself affect whether $v$ impacts $u$, introducing dependencies into the calculation of $\theta^{(x)}$ that would need to be handled carefully.

To avoid these dependencies, we will introduce the concept of a \emph{test node}~\cite{miller2011edge}.  The test-node assumption is closely related to the cavity-state used in message-passing models~\cite{karrer2010message}.  A single test node $u$ will be chosen randomly from the network.  We will perform a thought experiment, modifying the test node's interactions with its neighbors.  The test node's interactions with its neighbors proceed as if it were frozen in its initial status.  This makes the neighbors of $u$ independent of one another, regardless of whether $u$ has had a transition.  However, the status of node $u$ can undergo spontaneous transitions or neighbor-induced transitions exactly as any other node.  

The key observation is that until the first time $u$ changes its status, the test node concept has no effect.  By our assumption that no node can transition into a susceptible status, once $u$ changes status for the first time, it is no longer affected by its neighbors.  Thus the test node assumption has no impact on the status of $u$ at any time.  It can only impact the status of neighbors of $u$ (after $u$'s first transition).  The probability $u$ has a given status is the same with or without the test node assumption.  Since $u$ is randomly chosen from the population, the probability it has a given status at time $t$ should match the proportion of the population with that status at time $t$ (in the large network limit).

We define $\theta^{(x)}(t)$ to be the probability that as of time $t$, the random neighbor $v$ of the test node $u$ has not yet transmitted to the test node $u$.  With the test node assumption, the neighbors' statuses are independent of one another for all time, and so if $u$ has degree $k$, the probability $u$ has not received a transmission from any neighbor by time $t$ is $\theta^{(x)}(t)^k$.  

We now divide $\theta^{(x)}(t)$ into sub-compartments based on the status of the neighbor $v$.  We define $\phi_{z}^{(x)}$ to be the probability that a randomly chosen neighbor $v$ of the test node $u$ both has status $\Q_z$ at time $t$ and has not transmitted to $u$ prior to time $t$ under the assumption that the initial status of $u$ is $\Q_x$.  So $\theta^{(x)} = \sum_z \phi_z^{(x)}$.

\begin{figure}
\begin{tikzpicture}
    % A few representative subcompartments of theta^(x)
    \node[phi] (p1) at (-6.3,0) {$\phi_1^{(x)}$};
    \node[font=\large] at (-3.1,0) {$\cdots$};
    \node[phi] (py) at (0,0) {$\phi_y^{(x)}$};
    \node[font=\large] at (3.1,0) {$\cdots$};
    \node[phi] (pm) at (6.3,0) {$\phi_m^{(x)}$};

    \begin{scope}[on background layer]
      \node[thetafit,
            fit=(p1)(py)(pm),inner xsep=9mm,inner ysep=9mm]
            (thetaBox) {};
    \end{scope}
    \node[anchor=north west,font=\small,inner sep=0pt,
          xshift=3mm,yshift=-2mm] at (thetaBox.north west)
          {$\theta^{(x)}$};

    \node[1minustheta,minimum width=32mm] (T) at (0,-4.1)
          {$1-\theta^{(x)}$};

    % Transmission from each neighbor status to the test node
    \draw[flow] (p1.south) to[out=-80,in=175]
       node[rate,below, sloped ,pos=.46]
       {$\displaystyle\phi_1^{(x)}\sum_{z=\ell+1}^m b_{x,z}^{(1)}$} (T.west);
    \draw[flow] (py.south) -- node[rate,right,pos=.51]
       {$\displaystyle\phi_y^{(x)}\sum_{z=\ell+1}^m b_{x,z}^{(y)}$} (T.north);
    \draw[flow] (pm.south) to[out=-100,in=5]
       node[rate,below, sloped,pos=.46]
       {$\displaystyle\phi_m^{(x)}\sum_{z=\ell+1}^m b_{x,z}^{(m)}$} (T.east);
\end{tikzpicture}
\caption{Flow diagram showing $\theta^{(x)}$ divided into subcompartments of $\phi_y^{(x)}$ with edges to  $1-\theta^{(x)}$ having flux $\phi_{y}^{(x)} \sum_{z=\ell+1}^m b_{x,z}^{(y)}$ from each.  The sum $\sum_z b_{x,z}^{(y)}$ finds the rate at which a neighbor of status $\Q_y$ would induce a transition in a node of status $\Q_x$ into any possible status $\Q_z$.}
\label{fig:ExampleThetaFlow}
\end{figure}
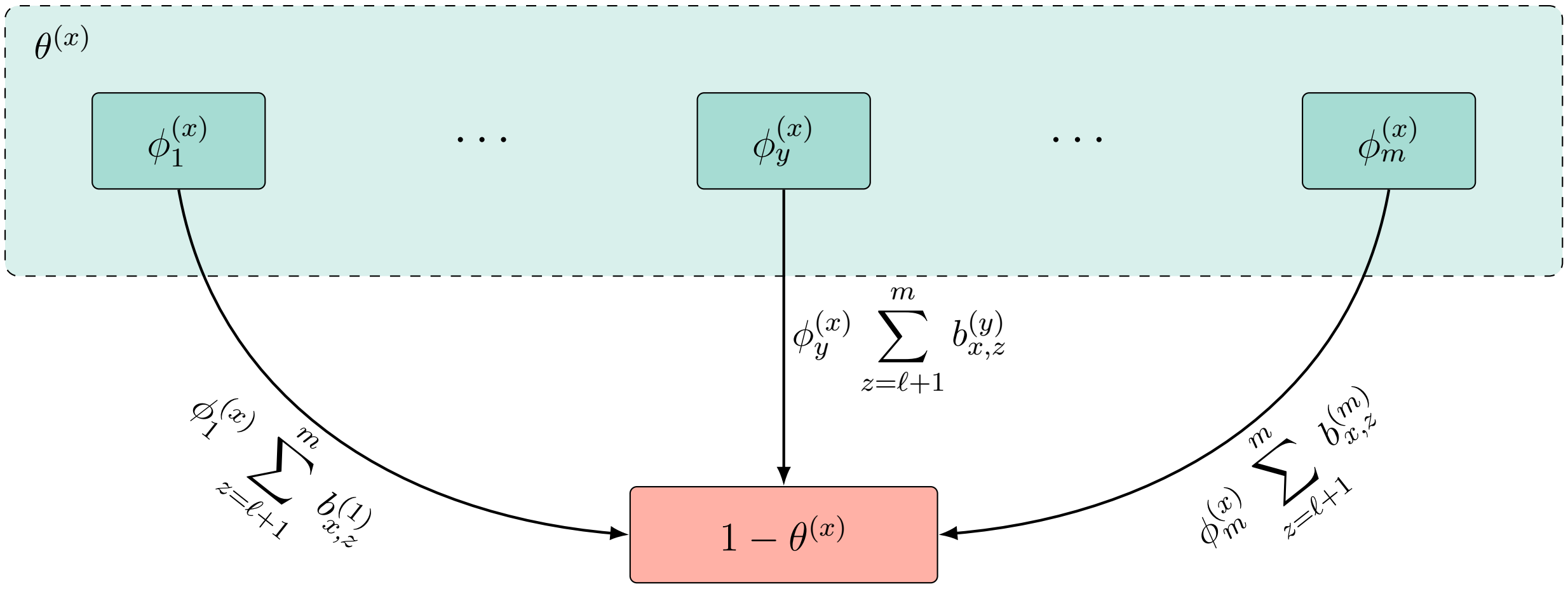

We can illustrate this with a partial flow diagram showing potential transmissions from the neighbor $v$ to the node $u$.  Following the standard approach in EBCM modeling we create a  compartment $\theta^{(x)}(t)$ and subdivide it into smaller compartments $\phi_y^{(x)}$.  There may be fluxes between the $\phi_y^{(x)}$ compartments corresponding to a neighbor changing status.  The flux out of one of these compartments due to transmission gives the rate of change of $\theta$ due to this type of transmission and we will use it to determine the flux from $\Q_x$ to a new (non-susceptible) status.  An illustrative flow diagram that does not show the flows between the $\phi_y^{(x)}$ compartments is provided in Fig.~\ref{fig:ExampleThetaFlow}.

We have
\begin{equation}
    \frac{d}{dt} \theta^{(x)} = - \sum_{y=1}^m \left(\phi_{y}^{(x)}\sum_{z=\ell+1}^m b_{x,z}^{(y)}\right) 
\end{equation}

\subsubsection{Flux out of Susceptible Statuses}
We know that if $\Q_x$ is a susceptible status, then $Q_x(t) = \rho_x e^{-t(\sum_z a_{x,z})} \psi(\theta^{(x)}(t))$.  Thus the time derivative is 
\begin{align*}
    \frac{d}{dt} Q_x(t) &= - \left(\sum_{z=\ell+1}^m a_{x,z}\right)Q_x(t)  + \rho_x e^{-t (\sum_{z=\ell+1}^m a_{x,z})} \psi'(\theta^{(x)}(t)) \frac{d}{dt} \theta^{(x)}(t)\\
    &= - \left(\sum_{z=\ell+1}^m a_{x,z}  \right)Q_x(t) - \rho_x e^{-t (\sum_{z=\ell+1}^m a_{x,z})} \psi'(\theta^{(x)}(t)) \sum_{y=1}^m \left(\phi_{y}^{(x)}(t)\sum_{z=\ell+1}^m b_{x,z}^{(y)}\right) \\
    &= - \sum_{z=\ell+1}^m \left( a_{x,z} Q_x(t) + \rho_x e^{-t (\sum_{\hat{z}=\ell+1}^m a_{x,\hat{z}})} \psi'(\theta^{(x)}(t)) \sum_{y=1}^m b_{x,z}^{y} \phi_{y}^{(x)}(t)  \right)
\end{align*}
By inspection, for each $z$, the term in the sum corresponds to the rate at which an individual of status $\Q_x$ transitions to status $\Q_z$ either due to a spontaneous transition (graph $\mathcal{H}$) or a transmission from a neighbor (graph $\mathcal{J}$).  Thus, the flux from $Q_x(t)$ to $Q_z(t)$ is $a_{x,z} Q_x(t) + \rho_x e^{-t \sum_{\hat{j}} a_{x,\hat{j}}} \psi'(\theta^{(x)}(t)) \sum_{y} b_{x,z}^{y} \phi_{y}^{(x)}(t)$.  Although this can be derived from first principles, we find that the simplest derivation comes through using the rules of derivatives as above and inferring the fluxes from corresponding parts of the resulting sum.

\subsubsection{Calculation of $\phi_{y}^{(x)}(t)$}

We can now calculate all of the $Q_x$ variables in terms of the $\theta$ and $\phi$ variables.  However, we need to complete our calculation of the $\phi$ variables.  This will proceed similarly to our calculations for statuses.

We now look to calculate $\phi_{y}^{(x)}(t)$.  We start by considering the case where $1 \leq y \leq \ell$, that is, where $\Q_y$ is a susceptible status.  We will find $\phi_{y}^{(x)}$ in terms of the $\theta$ variables.  From this we will be able to determine the flux from $\phi_y^{(x)}$ into the other $\phi$ variables (corresponding to the neighbor changing status) or into the $1-\theta^{(x)}$ compartment (corresponding to the neighbor transmitting).  

Before continuing, we note that $P_n(k) = kP(k)/\ave{K}$ is the probability that a random neighbor has degree $k$.  It accounts for the higher probability of connecting to a node with more neighbors~\cite{feld1991your,newman2001random,meyers2005network}.

For susceptible statuses $\Q_y$, we find (with $z$ taking values from $\ell+1$ to $m$)
\begin{align*}
\phi_y^{(x)} &= \rho_y e^{-t(\sum_z a_{y,z})} e^{-t (\sum_z b_{x,z}^{(y)})} e^{-t (\sum_z b_{y,z}^{(x)})} \sum_k P_n(k) \theta^{(y)}(t)^{k-1}\\
&= \rho_y e^{-t(\sum_z a_{y,z})} e^{-t (\sum_z b_{x,z}^{(y)})} e^{-t (\sum_z b_{y,z}^{(x)})} \frac{\psi'(\theta^{(y)}(t))}{\ave{K}}\\
&= \rho_y  \frac{e^{-t\sum_{z=\ell+1}^m (a_{y,z} + b_{x,z}^{(y)} + b_{y,z}^{(x)})}\psi'(\theta^{(y)}(t))}{\ave{K}}
\end{align*}
The $\rho_y$ gives the probability that the neighbor was initially of status $\Q_y$.  The first exponential corresponds to the probability that a neighbor of status $\Q_y$ will not undergo a spontaneous transition to some other status (assuming it receives no transmission from neighbors first).  The second exponential corresponds to the probability that the neighbor has not received a transmission from the test node (recall susceptible nodes may be able to transmit to susceptible neighbors).  The third exponential corresponds to the probability that the test node has not received a transmission from the neighbor.  The fraction involving $\psi'$ measures the probability that the neighbor has not received any transmission from any of its $k-1$ other neighbors.

Taking $\frac{d}{dt} \phi_y^{(x)}$, we arrive at
\begin{align*}
    \frac{d}{dt} \phi_y^{(x)} &= - \left(\sum_z (a_{y,z}  + b_{x,z}^{(y)} + b_{y, z}^{(x)} )\right)\phi_y^{(x)}\\
    &\quad + \rho_y\frac{e^{-t\sum_z (a_{y,z} + b_{x,z}^{(y)} + b_{y,z}^{(x)})}\psi''(\theta^{(y)}(t))}{\ave{K}} \frac{d}{dt} \theta^{(y)}(t)\\
    &= - \sum_z a_{y,z} \phi_y^{(x)} - \sum_z b_{x,z}^{(y)} \phi_y^{(x)} - \sum_z b_{y, z}^{(x)} \phi_y^{(x)}\\
    &\quad- \rho_y \frac{e^{-t\sum_{z=\ell+1}^m (a_{y,z} + b_{x,z}^{(y)} + b_{y,z}^{(x)})}\psi''(\theta^{(y)}(t))}{\ave{K}} \sum_{z=\ell+1}^m\sum_{v=1}^m b_{y,z}^{(v)} \phi_v^{(y)}\\
    &= - \sum_z b_{x,z}^{(y)} \phi_y^{(x)} \\
    &\quad- \sum_{z=\ell+1}^m \left(  a_{y,z} \phi_y^{(x)} +b_{y, z}^{(x)} \phi_y^{(x)} +\rho_y \frac{e^{-t \sum_{w=\ell+1}^m (a_{y,w} + b_{x,w}^{(y)} + b_{y,w}^{(x)})}\psi''(\theta^{(y)}(t))}{\ave{K}} \sum_{v=1}^m b_{y,z}^{(v)} \phi_v^{(y)}\right)
\end{align*}
The first term corresponds to the neighbor transmitting to the test node while still in a susceptible state.  This gives the flux from $\phi_y^{(x)}$ into $1-\theta^{(x)}$.  For each $z$, the remaining sum gives the flux from $\phi_y^{(x)}$ to $\phi_z^{(x)}$ due to the neighbor transitioning to status $\Q_z$.  This gives us the fluxes from $\phi_y^{(x)}$ for every susceptible status $\Q_y$.  From this, we find all the fluxes out of $\phi_y^{(x)}$ whenever $\Q_y$ is a susceptible status.  

The fluxes out of some $\phi_w^{(x)}$ where $\Q_w$ is not susceptible comes from the spontaneous transitions or from transmission to the test node.  The flux from $\phi_w^{(x)}$ to $\phi_z^{(x)}$ is $a_{w,z} \phi_w^{(x)}$.  The flux from $\phi_w^{(x)}$ to $1-\theta^{(x)}$ is $\sum_z b_{x,z}^{(w)} \phi_w^{(x)}$.  From these, we can identify all of the fluxes between the $\phi$ compartments in Fig.~\ref{fig:ExampleThetaFlow}.

\subsubsection{Full Equations}

We now know the fluxes between all of the variables.  Implicitly, this gives us a flow diagram.  We do not draw a general flow diagram for any such simple contagion here, but later where we consider specific examples, we give the corresponding flow diagrams.

Using the fluxes, we can write down a full system of equations for the simple contagion. 
We assume that the statuses $\Q_1, \ldots, \Q_\ell$ are susceptible and $\Q_{\ell+1}, \ldots, \Q_m$ are not.  Then for $x = 1, \ldots, \ell$ and $y= 1, \ldots, \ell$:
\begin{subequations}
\label{eqn:fullsystem}
\begin{align}
Q_x &= \rho_xe^{- t\sum_{z=\ell+1}^m a_{x,z}} \psi(\theta^{(x)}) \\
Q_x(0) &= \rho_x\\
\phi_{y}^{(x)} &=  \rho_y\frac{\psi'(\theta^{(y)})}{\ave{K}} e^{-t\sum_{z=\ell+1}^m \left(a_{y,z}+ b_{x,z}^{(y)} + b_{y,z}^{(x)}\right)} \\
\phi_y^{(x)}(0) &= \rho_y
\end{align}
For the non-susceptible statuses $\Q_z$ with $z = \ell+1, \ldots, m$, the fluxes in and out of $\Q_z$ give a differential equation for the probability a test node has status $\Q_z$,  Similarly, we can find fluxes in and out of $\phi_z^{(x)}$ to find the probability the test node's neighbor has status $\Q_z$ and has not transmitted to the test node given that the test node initially had susceptible status $\Q_x$.  For $x = 1, \ldots, \ell$, \ $z = \ell+1, \ldots, m$:
\begin{align}
  \frac{d}{dt} Q_z =& \sum_{x=1}^\ell \left( a_{x,z} Q_x(t) + \rho_x e^{-t (\sum_{w=\ell+1}^m a_{x,w})} \psi'(\theta^{(x)}) \sum_{y=1}^m b_{x,z}^{(y)} \phi_{y}^{(x)}(t)\right)\\
 &+ \sum_{w=\ell+1}^m \left(a_{w,z} Q_w(t) - a_{z,w} Q_z(t)\right)  \nonumber \\
 Q_z(0) =& \rho_z \\
\frac{d}{dt} \phi_z^{(x)} =& \sum_{y=1}^\ell \left( (a_{y,z} + b_{y,z}^{(x)}) \phi_y^{(x)}  + \frac{\rho_y e^{-t\sum_{w=\ell+1}^m \left(a_{y,w}+ b_{x,w}^{(y)} + b_{y,w}^{(x)}\right)} \psi''(\theta^{(y)})\sum_{v=1}^m \phi_v^{(y)} b_{y,z}^{(v)}}{\ave{K}}\right) \\
&+ \sum_{w=\ell+1}^m \left(a_{w,z}\phi_w^{(x)} - a_{z,w} \phi_{z}^{(x)}\right)- \sum_{w=\ell+1}^m b_{x,w}^{(z)}\phi_z^{(x)}  \nonumber \\
\phi_z^{(x)}(0) =& \rho_z 
\end{align}
Finally, we have an equation for $\frac{d}{dt}\theta^{(x)}$, the probability the neighbor has not transmitted to the test node.
\begin{align}
\frac{d}{dt} \theta^{(x)} =& - \sum_{y=1}^m \sum_{z=\ell+1}^m b_{x,z}^{(y)} \phi_{y}^{(x)} \\
\theta^{(x)}(0) =& 1
\end{align}
\end{subequations}
Note that we could replace the equation for $\frac{d}{dt} \theta^{(x)}(t)$ with $\theta^{(x)}(t) = \sum_{y=1}^m \phi_y^{(x)}$.

We reiterate that these equations apply when the simple contagion has the property that it is not possible to enter a susceptible status.  Nodes may begin with a susceptible status, but if their status ever changes, they are never again susceptible.

In deriving the expression above, we assumed that the susceptible statuses are indexed from $1$ up to $\ell$.  If the statuses are not ordered in this way then we would replace the sums over $1$ to $\ell$ with sums over the susceptible statuses.  Similarly we would replace sums from $1$ to $m$ with sums over all statuses and sums from $\ell+1$ to $m$ with sums over all non-susceptible statuses.

It is possible to automate the process of solving these equations starting from the graphs $\mathcal{H}$ and $\mathcal{J}$, the degree distribution, and the values $\rho_y$.  This is done in version 2.1 of the Epidemics on Networks package~\cite{miller2019EoN}\footnote{This version is currently in final development and will be released in Dec 2026.  The development version can be accessed at {\color{blue}\url{https://github.com/springer-math/Mathematics-of-Epidemics-on-Networks}}}, based on Networkx~\cite{Hagberg2008networkx}.

\section{Examples}

We consider the three examples from Fig.~\ref{fig:simple_contagion_HJ}.  

\subsection{SEIR Model}
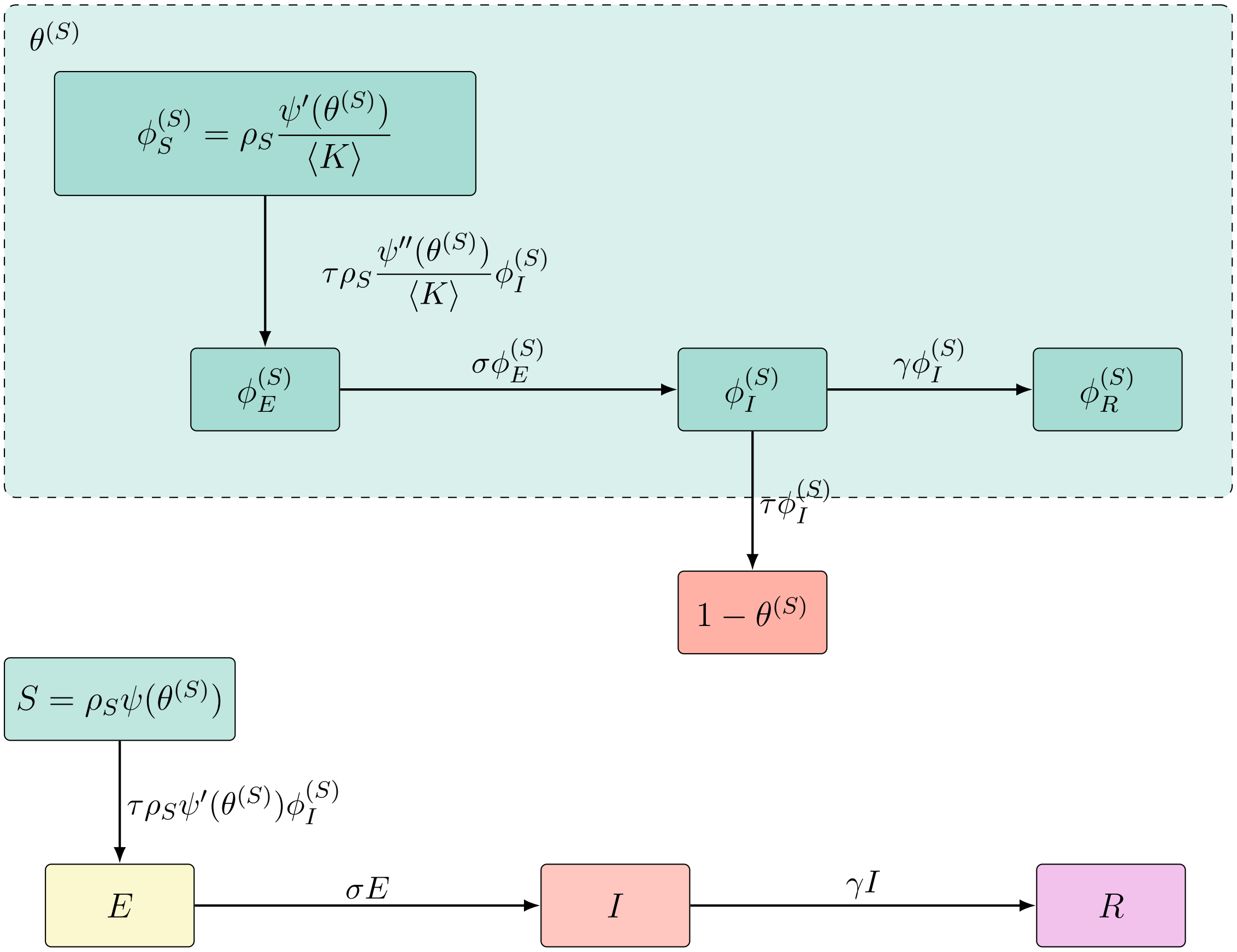
\begin{figure}
\begin{tikzpicture}
  % The susceptible neighbor state is directly above the exposed one.
  \node[phi,minimum width=51mm,minimum height=15mm] (S) at (0,0)
    {$\displaystyle \phi_S^{(S)}=\rho_S\frac{\psi'(\theta^{(S)})}{\langle K\rangle}$};
  \node[phi] (E) at (0,-3.1) {$\phi_E^{(S)}$};
  \node[phi] (I) at (5.9,-3.1) {$\phi_I^{(S)}$};
  \node[phi] (R) at (10.2,-3.1) {$\phi_R^{(S)}$};
  \node[1minustheta] (T) at (5.9,-5.8) {$1-\theta^{(S)}$};

  \begin{scope}[on background layer]
    \node[thetafit,fit=(S)(E)(I)(R),
          inner xsep=6mm,inner ysep=8mm] (thetaBox) {};
  \end{scope}
  % Place theta inside the upper-left of the dashed enclosure.
  \node[anchor=north west,font=\small,inner sep=0pt,
        xshift=3mm,yshift=-2mm] at (thetaBox.north west)
        {$\theta^{(S)}$};

  \draw[flow] (S.south) -- node[rate,right,xshift=6mm]
    {$\displaystyle \tau\rho_S
      \frac{\psi''(\theta^{(S)})}{\langle K\rangle}
      \phi_I^{(S)}$} (E.north);
  \draw[flow] (E.east) -- node[rate,above] {$\sigma\phi_E^{(S)}$} (I.west);
  \draw[flow] (I.east) -- node[rate,above] {$\gamma\phi_I^{(S)}$} (R.west);
  \draw[flow] (I.south) -- node[rate,right] {$\tau\phi_I^{(S)}$} (T.north);
\end{tikzpicture}

\begin{tikzpicture}
    \node[Sstatus] (S) at (0,0) {$S=\rho_S\psi(\theta^{(S)})$};
    \node[Estatus] (E) at (0,-2.5) {$E$};
    \node[Istatus] (I) at (6,-2.5) {$I$};
    \node[Rstatus] (R) at (12,-2.5) {$R$};

    \draw[flow] (S) -- node[rate,right]
        {$\tau\rho_S\psi'(\theta^{(S)})\phi_I^{(S)}$} (E);

    \draw[flow] (E) -- node[rate,above]
        {$\sigma E$} (I);

    \draw[flow] (I) -- node[rate,above]
        {$\gamma I$} (R);
\end{tikzpicture}

\caption{Flow diagrams corresponding to the SEIR model in Fig.~\ref{fig:simple_contagion_HJ}(a).  \textbf{Top}: Flow diagram for $\phi^{(S)}$ and $\theta^{(S)}$ variables corresponding to an initially susceptible test node.  \textbf{Bottom}: Flow diagram for the proportion of nodes with each status.} 
\label{fig:SEIR}
\end{figure}

The first is a well-known SEIR model whose transition graphs are shown in Fig.~\ref{fig:simple_contagion_HJ}(a).  The flow diagram is shown in Fig.~\ref{fig:SEIR}.  By taking $\Q_1=\Sus$, \ $\Q_2=\E$, \ $\Q_3=\I$, and $\Q_4=\R$ our earlier equations apply to this scenario.  In this case there are $\ell=1$ susceptible statuses and $m=4$ total statuses.  The EBCM equations for this model also follow from equations for multi-stage infection models previously considered in~\cite{miller2013incorporating}.

The system of equations is
\begin{align*}
    S &= \rho_S \psi(\theta^{(S)})\\
    \frac{d}{dt} E &= \tau \rho_S \psi'(\theta^{(S)})\phi_I^{(S)} - \sigma E\\
    \frac{d}{dt} I &= \sigma E - \gamma I\\
    \frac{d}{dt} R &= \gamma I\\
    \phi_S^{(S)} &=  \rho_S\frac{\psi'(\theta^{(S)})}{\ave{K}}\\
    \frac{d}{dt} \phi_E^{(S)} &= \tau \rho_S \frac{\psi''(\theta^{(S)})}{\ave{K}} \phi_I^{(S)} - \sigma \phi_E^{(S)}\\
    \frac{d}{dt} \phi_I^{(S)} &= \sigma \phi_E^{(S)} - \gamma \phi_I^{(S)} -\tau \phi_I^{(S)}\\
    \frac{d}{dt} \phi_R^{(S)} &= \gamma \phi_I^{(S)}\\
    \frac{d}{dt} \theta^{(S)} &= -\tau \phi_I^S
\end{align*}

\begin{figure}
    \includegraphics{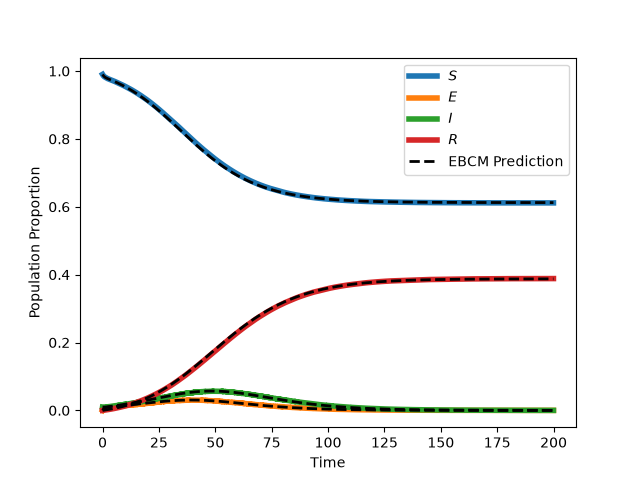}
    \caption{Illustrative dynamics of the SEIR model of Fig.~\ref{fig:simple_contagion_HJ}(a) on a network comparing stochastic simulations and solutions to the EBCM model.  Parameters are $\tau=1$, \ $\sigma = 0.2$, \ $\gamma=0.1$. The initial conditions are $S(0)=0.99$, \ $I(0)=0.01$ and $E(0)=R(0)=0$.  The network has a negative binomial degree distribution with $p=0.2$ and $r=5$, corresponding to $\psi(x) = \left(\frac{1-p}{1-px}\right)^r$ with $N=10^6$ nodes.}
    \label{fig:SEIR_comp}
\end{figure}

To illustrate the accuracy of the EBCM equations, we compare simulations with the deterministic prediction in Fig.~\ref{fig:SEIR_comp} in a network of $10^6$ nodes having a negative binomial degree distribution.  The agreement is very good.

\subsection{A model with spontaneous status change for susceptible nodes}
\begin{figure}
    \begin{tikzpicture}
  % Middle model in Fig. 2: spontaneous S -> I_1 and recovery.
  % For compactness define the remaining-neighbor flux factor F(t).
  \node[phi,minimum width=53mm,minimum height=15mm] (S) at (0,0)
       {$\displaystyle \phi_S^{(S)}=\rho_S e^{-\omega t}
            \frac{\psi'(\theta^{(S)})}{\langle K\rangle}$};
  \node[phi] (Ione) at (0,-3.5) {$\phi_{I_1}^{(S)}$};
  \node[phi] (R1) at (4,-3.5) {$\phi_{R_1}^{(S)}$};
  \node[phi] (Itwo) at (8.0,-3.5) {$\phi_{I_2}^{(S)}$};
  \node[phi] (R2) at (12,-3.5) {$\phi_{R_2}^{(S)}$};
  \node[1minustheta] (T) at (5.0,-7.0) {$1-\theta^{(S)}$};

  \begin{scope}[on background layer]
    \node[thetafit,fit=(S)(Ione)(Itwo)(R1)(R2),
          inner xsep=7mm,inner ysep=8mm] (thetaBox) {};
  \end{scope}
  \node[anchor=north west,font=\small,inner sep=0pt,
        xshift=3mm,yshift=-2mm] at (thetaBox.north west)
        {$\theta^{(S)}$};

  \draw[flow] (Ione) --       node[rate,above] {$\gamma_1\phi_{I_1}^{(S)}$}
       (R1);
  \draw[flow] (Itwo.east) --
       node[rate,above] {$\gamma_2\phi_{I_2}^{(S)}$} (R2.west);
  \draw[flow] (S.south) --
      node[left,xshift=1.5mm]
        {$\omega\phi_S^{(S)}\!\!+\!\!p\tau_1\phi_{I_1}^{(S)}F(t)$} (Ione.north);
  \draw[flow]
       (S.east) -- node[sloped, above] 
       {$F(t)\left[(1-p)\tau_1\phi_{I_1}^{(S)}+\tau_2\phi_{I_2}^{(S)}\right]$} (Itwo.north);

  % Transmission to test node: two distinct hazards.
  \draw[flow] (Ione.south)   --     node[rate,below,sloped] {$\tau_1\phi_{I_1}^{(S)}$}
       (T.north);
  \draw[flow] (Itwo.south) --
       node[rate,below,sloped] {$\tau_2\phi_{I_2}^{(S)}$} (T.north);

  % Identity used to keep the arrow labels short.

    \node[Sstatus] (S)  at (-1,-10.5) {$S=\rho_Se^{-\omega t} \psi(\theta^{(S)})$};
    \node[Istatus] (I1) at (8,-9) {$I_1$};
    \node[Istatus] (I2) at (8,-12) {$I_2$};
    \node[Rstatus] (R1)  at (11,-9) {$R_1$};
    \node[Rstatus] (R2)  at (11,-12) {$R_2$};

    \draw[flow] (S) -- node[rate,sloped,above]
        {$\omega S + p\tau_1 G(t)\phi_{I_1}^{(S)}$} (I1);

    \draw[flow] (S) -- node[rate,sloped,below]
        {$G(t)\!\left[(1-p)\tau_1\phi_{I_1}^{(S)}+\tau_2\phi_{I_2}^{(S)}\right]$} (I2);

    \draw[flow] (I1) -- node[rate,sloped,above]
        {$\gamma_1 I_1$} (R1);

    \draw[flow] (I2) -- node[rate,sloped,below]
        {$\gamma_2 I_2$} (R2);

        \node[font=\small,anchor=north,align=center] at (3.8,-13)
       {$\displaystyle F(t)=\rho_S e^{-\omega t}
          \frac{\psi''(\theta^{(S)})}{\langle K\rangle} \qquad \qquad G(t)=\rho_S e^{-\omega t}\psi'(\theta^{(S)})$};

\end{tikzpicture}

\caption{Flow diagrams corresponding to Fig.~\ref{fig:simple_contagion_HJ}(b). \textbf{Top}: Flow diagram for the $\phi^{(S)}$ and $\theta^{(S)}$ variables corresponding to an initially susceptible test node.  \textbf{Bottom}: Flow diagram for the proportion of nodes with each status.}
\label{fig:SI1I2RFlow}
\end{figure}
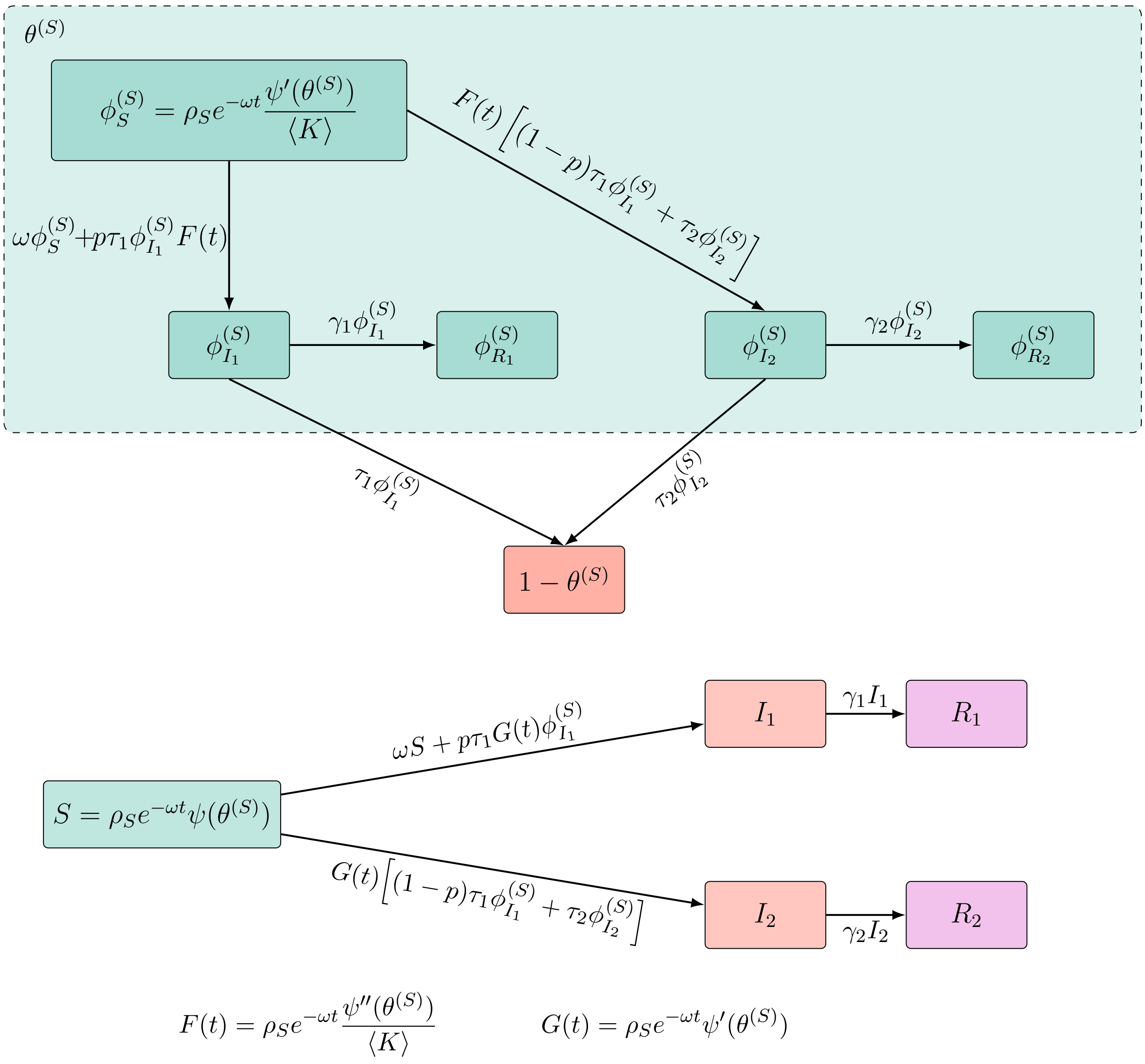

In the second model, susceptible individuals can spontaneously become infected.  There are two different types of infection that can occur.  The transition graphs are shown in Fig.~\ref{fig:simple_contagion_HJ}(b).  The flow diagram is shown in Fig.~\ref{fig:SI1I2RFlow}.

\begin{figure}
    \includegraphics{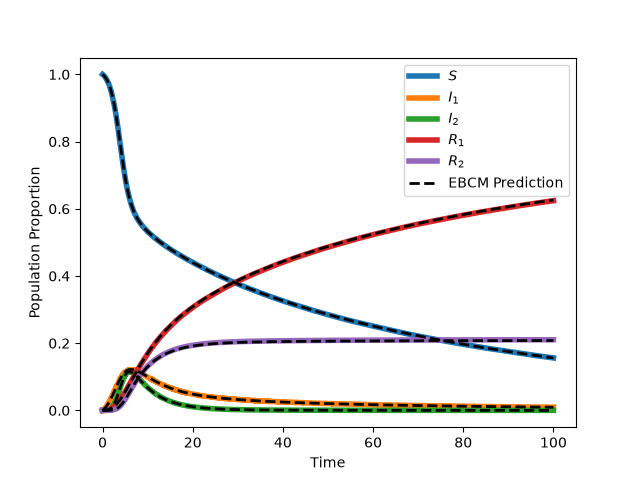}
    \caption{Illustrative dynamics of the SI$_1$I$_2$R$_1$R$_2$ model of Fig.~\ref{fig:simple_contagion_HJ}(b) on a network comparing stochastic simulations and solutions to the EBCM model.  Parameters are $\omega = 0.01$, \ $\gamma_1 = 0.2$, \ $\gamma_2 = 0.2$, \ $\tau_1 = 1$, \ $\tau_2 = 3$, and $p = 0.95$. The initial conditions are $S(0)=1$ and $I_1(0)=I_2(0)=R_1(0)=R_2(0)=0$. The network has the same properties as in Fig.~\ref{fig:SEIR_comp}.}
    \label{fig:SI1I2R_comp}
\end{figure}

The equations are
\begin{align*}
    S &= \rho_S e^{-\omega t} \psi(\theta^{(S)})\\
    \frac{d}{dt} I_1 &= \omega S + p \tau_1 G(t) \phi_{I_1}^{(S)} - \gamma_1I_1\\
    \frac{d}{dt} I_2 &= G(t)\left[(1-p) \tau_1 \phi_{I_1}^{(S)}+ \tau_2 \phi_{I_2}^{(S)}\right] - \gamma_2 I_2\\
    \frac{d}{dt} R_1 &= \gamma_1 I_1\\
    \frac{d}{dt} R_2 &=\gamma_2 I_2\\
    \phi_S^{(S)} &= \rho_S e^{-\omega t} \frac{\psi'(\theta^{(S)})}{\ave{K}}\\
    \frac{d}{dt} \phi_{I_1}^{(S)} &= \omega \phi_S^{(S)} + p \tau_1 \phi_{I_1}^{(S)}F(t) - (\gamma_1 + \tau_1)\phi_{I_1}^{(S)}\\
    \frac{d}{dt} \phi_{I_2}^{(S)} &= F(t)\left[(1-p)\tau_1\phi_{I_1}^{(S)}+\tau_2\phi_{I_2}^{(S)}\right] - (\gamma_2+\tau_2)\phi_{I_2}^{(S)}\\
    \frac{d}{dt} \phi_{R_1}^{(S)} &= \gamma_1 \phi_{I_1}^{(S)}\\
    \frac{d}{dt} \phi_{R_2}^{(S)} &= \gamma_2 \phi_{I_2}^{(S)}\\
    \frac{d}{dt}\theta^{(S)} &= -\tau_1 \phi_{I_1}^{(S)} - \tau_2 \phi_{I_2}^{(S)}\\
    F(t)&=\rho_S e^{-\omega t}
          \frac{\psi''(\theta^{(S)})}{\langle K\rangle} \\
    G(t)&=\rho_S e^{-\omega t}\psi'(\theta^{(S)})
\end{align*}
To illustrate the accuracy of the EBCM equations, we compare the solutions with stochastic simulations in Fig.~\ref{fig:SI1I2R_comp}.  Again the fit is very good.

\subsection{A model with susceptible statuses that are also infectious}
Finally we consider the third model from Fig.~\ref{fig:simple_contagion_HJ}.  There are two types of susceptible node, and each can cause other susceptible nodes to change status.  The flow diagram is given in Fig.~\ref{fig:S1S2I1I2RFlow}.
\begin{figure}
    \begin{tikzpicture}
  % Test node is frozen in S_1.  The other theta-system supplies A_2.
  \node[phi,minimum width=50mm,minimum height=16mm] (phiS1) at (-0.5,1)
    {$\displaystyle\phi_{S_1}^{(S_1)}=
       \rho_{S_1}\frac{\psi'(\theta^{(S_1)})}{\langle K\rangle}$};
  \node[phi,minimum width=61mm,minimum height=16mm] (phiS2) at (-1,-3)
    {$\displaystyle\phi_{S_2}^{(S_1)}=
       \rho_{S_2}e^{-(\alpha+\beta)t}
       \frac{\psi'(\theta^{(S_2)})}{\langle K\rangle}$};
  \node[phi] (phiI1) at (8,1) {$\phi_{I_1}^{(S_1)}$};
  \node[phi] (phiI2) at (8,-3.0) {$\phi_{I_2}^{(S_1)}$};
  \node[phi] (phiR) at (12,-1) {$\phi_R^{(S_1)}$};
  \node[1minustheta,minimum width=31mm] (T) at (5,-7.0)
    {$1-\theta^{(S_1)}$};

  \begin{scope}[on background layer]
    \node[thetafit,
          fit=(phiS1)(phiS2)(phiI1)(phiI2)(phiR),
          inner xsep=8mm,inner ysep=10mm] (thetaBox) {};
  \end{scope}
  \node[anchor=north west,font=\small,inner sep=0pt,
        xshift=3mm,yshift=-2mm] at (thetaBox.north west)
    {$\theta^{(S_1)}$};

  % The neighbor's susceptible-to-infectious transitions.
  \draw[flow] (phiS1) --
      node[rate,above,xshift=-1.5mm] {$F_1(t)A_{11}(t)$}%{$F_1(t)A_1(t)$} 
      (phiI1);
  \draw[flow] (phiS1) -- node[rate, above,sloped, pos=0.65] {$F_1(t)A_{12}(t)$} (phiI2);
  \draw[flow] (phiS2) -- node[rate, above, sloped, pos=0.65] {$C(t)F_2(t)A_{21}(t)$} (phiI1);
  \draw[flow] (phiS2) -- node[rate, above, pos = 0.45]
      %{$\alpha\phi_{S_2}^{(S_1)}$\\$+F_2(t)A_2(t)$} 
      {$\alpha \phi_{S_2}^{(S_1)} + C(t)F_2(t)A_{22}(t)$} (phiI2);

  % Recoveries.  The first arrow arches above I_2.
  \draw[flow]
       (phiI1) -- node[rate,sloped,above] {$\gamma_1\phi_{I_1}^{(S_1)}$}
       (phiR);
  \draw[flow] (phiI2) --
       node[rate,sloped,above] {$\gamma_2\phi_{I_2}^{(S_1)}$} (phiR);

  % Transmissions from the neighbor to the S_1 test node.
  \draw[flow] (phiI1) --
      node[rate,above,sloped, pos=0.7] {$\tau_{11}\phi_{I_1}^{(S_1)}$}
      (T);
  \draw[flow] (phiI2) --
      node[rate,below,sloped] {$\tau_{21}\phi_{I_2}^{(S_1)}$} (T);
  \draw[flow] (phiS2) -- node[rate,above,sloped] {$\beta\phi_{S_2}^{(S_1)}$}
      (T);

  % Abbreviations for transitions induced by other neighbors.
  
    \node[Sstatus] (S1) at (0,-9.5) {$S_1=\rho_{S_1}\psi(\theta^{(S_1)})$};
    \node[Sstatus] (S2) at (0,-12.5) {$S_2=\rho_{S_2}\psi(\theta^{(S_2)})$};
    \node[Istatus] (I1) at (8,-9.5) {$I_1$};
    \node[Istatus] (I2) at (8,-12.5) {$I_2$};
    \node[Rstatus] (R) at (12,-11) {$R$};

    % Draw diagonals first; labels are placed away from the crossing.
    \draw[flow] (S1) -- node[rate,sloped,above,pos=.29]
        {$G_1\tau_{21}\phi_{I_2}^{(S_1)}$} (I2);
    \draw[flow] (S2) -- node[rate,sloped,below,pos=.71]
        {$G_2\tau_{12}\phi_{I_1}^{(S_2)}$} (I1);

    \draw[flow] (S1) -- node[rate,above]
        {$G_1\bigl(\tau_{11}\phi_{I_1}^{(S_1)}+\beta\phi_{S_2}^{(S_1)}\bigr)$} (I1);
    \draw[flow] (S2) -- node[rate,below]
        {$G_2\bigl(\tau_{22}\phi_{I_2}^{(S_2)}+\alpha\phi_{S_1}^{(S_2)}\bigr)$} (I2);

    \draw[flow] (I1) -- node[rate,sloped,above]
        {$\gamma_1 I_1$} (R);
    \draw[flow] (I2) -- node[rate,sloped,below]
        {$\gamma_2 I_2$} (R);

    % Definitions below the figure, without boxes.
    \node[anchor=north,align=left,font=\small] at (5,-15) {
    $\displaystyle F_1(t)=\rho_{S_1}\frac{\psi''(\theta^{(S_1)})}{\langle K\rangle},\qquad
       F_2(t)=\rho_{S_2}\frac{\psi''(\theta^{(S_2)})}{\langle K\rangle}, \qquad C(t)=e^{-(\alpha+\beta)t}$\\[6pt]
    $\displaystyle A_{11}(t)=\beta\phi_{S_2}^{(S_1)}+\tau_{11}\phi_{I_1}^{(S_1)},\qquad
       A_{12}(t)=\tau_{21}\phi_{I_2}^{(S_1)} \qquad  A_{21}(t)=\tau_{12}\phi_{I_1}^{(S_2)},\qquad
       A_{22}(t)=\alpha\phi_{S_1}^{(S_2)}+\tau_{22}\phi_{I_2}^{(S_2)}$\\[6pt]
    $\displaystyle G_1=\rho_{S_1}\psi'\bigl(\theta^{(S_1)}\bigr), \qquad G_2=\rho_{S_2}\psi'\bigl(\theta^{(S_2)}\bigr)$
  };
  
\end{tikzpicture}

\caption{Flow diagrams corresponding to Fig.~\ref{fig:simple_contagion_HJ}(c).  \textbf{Top}: Flow diagram for the $\phi^{(S_1)}$ and $\theta^{(S_1)}$ variables corresponding to a test node of initial status $\Sus_1$.  The diagram for $\phi^{(S_2)}$ and $\theta^{(S_2)}$ variables is symmetric. \textbf{Bottom}: Flow diagram for the proportion of nodes with each status.}
\label{fig:S1S2I1I2RFlow}
\end{figure}

\begin{figure}
    \includegraphics{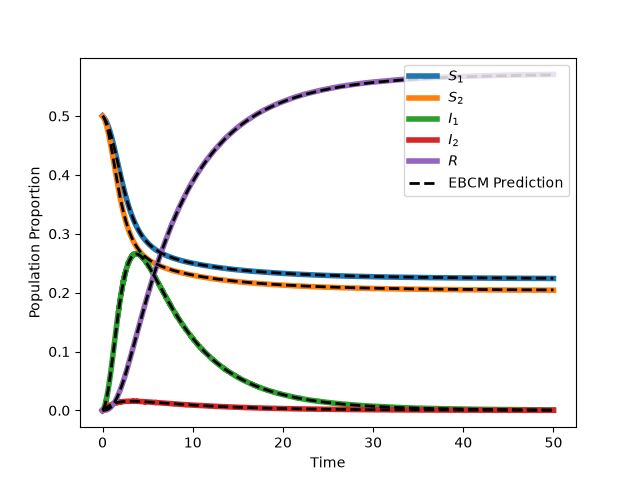}
    \caption{Illustrative dynamics of the S$_1$S$_2$I$_1$I$_2$R model of Fig.~\ref{fig:simple_contagion_HJ}(c) on a network comparing stochastic simulations and solutions to the EBCM model.  Parameters are $\alpha=0.04$, \ $\beta = 0.05$, \ $\tau_{11}=1$, \ $\tau_{12} = 3$, \ $\tau_{21}=1$, \ $\tau_{22}=3$, \ $\gamma_1=0.2$, and $\gamma_2=0.2$. The initial conditions are $S_1(0)=S_2(0)=1/2$ and $I_1(0)=I_2(0)=R=0$. The network has the same properties as in Fig.~\ref{fig:SEIR_comp}.}
    \label{fig:S1S2I1I2R_comp}
\end{figure}

The resulting equations are

\begin{align*}
S_1 &= \rho_{S_1}\psi(\theta^{(S_1)}), &
S_2 &= \rho_{S_2}\psi(\theta^{(S_2)}),\\
\frac{d}{dt} I_1 &= G_1 A_{11}+G_2 A_{21}-\gamma_1I_1, &
\frac{d}{dt} I_2 &= G_1 A_{12}+G_2 A_{22}-\gamma_2I_2,\\
\frac{d}{dt} R &= \gamma_1I_1+\gamma_2I_2,\\
\phi_{S_1}^{(S_1)} &= \rho_{S_1}\frac{\psi'(\theta^{(S_1)})}{\langle K\rangle},&
\phi_{S_2}^{(S_2)} &= \rho_{S_2}\frac{\psi'(\theta^{(S_2)})}{\langle K\rangle},\\
\phi_{S_2}^{(S_1)} &= C\rho_{S_2}\frac{\psi'(\theta^{(S_2)})}{\langle K\rangle},&
\phi_{S_1}^{(S_2)} &= C\rho_{S_1}\frac{\psi'(\theta^{(S_1)})}{\langle K\rangle},\\
\frac{d}{dt}\phi_{I_1}^{(S_1)} &= F_1 A_{11}+CF_2 A_{21}-(\gamma_1+\tau_{11})\phi_{I_1}^{(S_1)},&
\frac{d}{dt}\phi_{I_2}^{(S_2)} &= CF_1 A_{12}+F_2 A_{22}-(\gamma_2+\tau_{22})\phi_{I_2}^{(S_2)},\\
\frac{d}{dt}\phi_{I_2}^{(S_1)} &= F_1 A_{12}+\alpha\phi_{S_2}^{(S_1)}+CF_2 A_{22}
&\frac{d}{dt}\phi_{I_1}^{(S_2)} &= \beta\phi_{S_1}^{(S_2)}+CF_1 A_{11}+F_2 A_{21}\\
&\quad -(\gamma_2+\tau_{21})\phi_{I_2}^{(S_1)}, &&
\quad -(\gamma_1+\tau_{12})\phi_{I_1}^{(S_2)},\\
\frac{d}{dt}\phi_R^{(S_1)} &= \gamma_1\phi_{I_1}^{(S_1)}+\gamma_2\phi_{I_2}^{(S_1)},&
\frac{d}{dt}\phi_R^{(S_2)} &= \gamma_1\phi_{I_1}^{(S_2)}+\gamma_2\phi_{I_2}^{(S_2)},\\
\frac{d}{dt}\theta^{(S_1)} &= -A_{11}-A_{12}, &
\frac{d}{dt}\theta^{(S_2)} &= -A_{21}-A_{22}
\end{align*}
with
\begin{gather*}
A_{11} = \tau_{11}\phi_{I_1}^{(S_1)}+\beta\phi_{S_2}^{(S_1)},\qquad
A_{12} = \tau_{21}\phi_{I_2}^{(S_1)},\qquad
A_{21} = \tau_{12}\phi_{I_1}^{(S_2)},\qquad
A_{22} = \tau_{22}\phi_{I_2}^{(S_2)}+\alpha\phi_{S_1}^{(S_2)},\\
C(t) = e^{-(\alpha+\beta)t},\\
F_i(t) = \rho_{S_i}\frac{\psi''(\theta^{(S_i)})}{\langle K\rangle}, \qquad
G_i(t) = \rho_{S_i}\psi'(\theta^{(S_i)}),\qquad i=1,2,\\
\theta^{(S_i)}(0) = 1,\quad
\phi_v^{(S_i)}(0)=
Q_v(0)=\rho_v,\qquad i=1,2  \text{ and for all statuses } \Q_v
\end{gather*}
To illustrate the accuracy of the EBCM equations, we compare the solutions with stochastic simulations in Fig.~\ref{fig:S1S2I1I2R_comp}.  Once again the fit is very good.

\section{Discussion}

We have derived an EBCM model for a broad class of simple contagions.  The simple contagions are constrained so that a node can never transition into a susceptible status and we require that the initial condition of each node is assigned independently of the other nodes.  The resulting EBCM equations can be derived and solved automatically so that we only need to know the possible node transitions, the distribution of initial conditions, and the degree distribution of the network.  An algorithm that does this is provided in version 2.1 of the \texttt{Epidemics on Networks} package~\cite{miller2019EoN}.

We find that the solutions to the EBCM equations accurately predict the large-population limit of stochastic simulations.  We note that we have not provided a proof that these equations are the correct limit.  It is more accurate to say that our argument shows that if there is a large-population deterministic limit to the stochastic solutions, it will follow the EBCM equations we have derived.  

We highlight two specific cases where we expect the large-population limit to not converge to a deterministic limit:
\begin{itemize}
    \item Networks in which the degree distribution has a long tail.  For example in the SIR model, the highest degree nodes become infected soon (with a rate proportional to their degree) and then contribute more infections (with a rate proportional to their degree).  Thus the impact of a node at early time is proportional to the square of its degree.  If the tail of the distribution decays slowly, then as we generate larger and larger networks, the square of the degree of the highest degree node grows as quickly (or faster than) the number of nodes in the network.  Thus a single node may have an impact at the scale of the entire population --- the specific time of its infection or the duration of its infection are stochastic variables that have a non-negligible impact on the dynamics.  If so, it is not possible to write down a deterministic model that can capture this.  We believe in some cases (such as the SIR model) once the highest degree nodes are no longer infectious, an EBCM model would be accurate for the remaining residual network, but we have not addressed this here.
    \item Scenarios in which the initial number of nodes in some status remains $0$ or $\mathcal{O}(1)$ as $N \to \infty$.  This is discussed in the SIR case by~\cite{janson2014law}.  In the SEIR model used here, if we begin with $E(0)=0$ and $I(0)=1$, then the later timing of any epidemic depends on the early stochastic dynamics while the number of infections grows (or fails to grow) large enough to behave deterministically.  Our examples show cases in which we begin with $0$ nodes in some statuses.  So we do not claim that EBCM equations are never appropriate if some statuses may not be present initially.  We do not address the specifics in detail here, but note that it depends on whether stochastic effects can dramatically impact whether, or how quickly, counts begin to grow in different simulations.  Again, we anticipate that once counts are large enough that stochastic effects become insignificant on the population scale, then the EBCM model will apply on the network.
\end{itemize}

Although we have identified a class of simple contagions for which EBCM equations exist, this is not a complete categorization of all such simple contagions.  Others do exist. For example, EBCM equations can be found for contagions in which susceptible nodes transition between susceptible statuses as long as those susceptible statuses are equally susceptible to receiving a transmission from any given infectious state.

\bibliographystyle{plain}

\bibliography{EBCM_SimpleContagion}
\end{document}